\documentclass[pra,aps,amsfonts,amsmath,superscriptaddress,%
  twocolumn,showpacs,floatfix]{revtex4}
\usepackage{graphicx,latexsym,amssymb,amsmath,color, multirow, mathrsfs, booktabs}
\usepackage{amsfonts}
\usepackage{bm}
\usepackage{epsfig}

\begin{document}
\title{Giant neutron halos in the non-relativistic mean field approach}

\author{M. Grasso}
\affiliation{Institut de Physique Nucl\'eaire, 15 rue Georges
  Cl\'emenceau, F-91406 Orsay Cedex, France}
\affiliation{Dipartimento di Fisica e Astronomia, Via Santa
  Sofia 64, I-95123 Catania, Italy}
\affiliation{INFN, Sezione di Catania, Via Santa
  Sofia 64, I-95123 Catania, Italy}
\author{S. Yoshida}
\affiliation{Institut de Physique Nucl\'eaire, 15 rue Georges
  Cl\'emenceau, F-91406 Orsay Cedex, France}
\affiliation{Science Research Center, Hosei University, 2-17-1 Fujimi,
Chiyoda, Tokyo 102-8160, Japan}
\author{N. Sandulescu}
\affiliation{Service de Physique Nucl\'eaire, CEA-DAM Ile-de-France, BP 12,
F-91680 Bruy\`eres-le-Ch\^atel, France}
\affiliation{Institute for Physics and Nuclear Engineering, P.O. Box MG-6,
76900 Bucharest, Romania}
\author{N. Van Giai}
\affiliation{Institut de Physique Nucl\'eaire, 15 rue Georges
  Cl\'emenceau, F-91406 Orsay Cedex, France}

\begin{abstract}
Giant neutron halos in medium-heavy nuclei are studied in the
framework of the Hartree-Fock-Bogoliubov (HFB) approach with Skyrme
interactions. The appearance of such structures depends sensitively
on the effective interaction adopted. This is illustrated by
comparing the predictions of SLy4 and SkI4 in the Ca and Zr isotopic
chains. The former force gives no halo effect, the latter predicts a
neutron halo in the Zr chain with A$>$122 due to the weakly bound
orbitals 3$p1/2$ and 3$p3/2$. The structure of the
halo is analyzed in terms of the occupation probabilities of these
orbitals and their partial contributions to the neutron density. The
anti-halo effect in Ni and Zr isotopes is also discussed by
comparing the occupation probabilities of Hartree-Fock neutron
single-particle states near the Fermi energy with the corresponding
HFB quasiparticle states.
\end{abstract}
\maketitle

\section{Introduction}

In the present years several projects for the construction of a new
generation of radioactive beam facilities are in progress (see,
e.g.,Ref. \cite{pro}). Such facilities will permit to investigate
the properties of unstable nuclei situated close to the drip line
regions. From the theoretical side, many efforts are devoted to
perform accurate predictions to locate the proton and neutron drip
lines as well as to describe the behavior of unstable nuclei.
Unfortunately, theoretical predictions in exotic regions of the
nuclear chart can be rather model dependent.

Self-consistent mean field methods are well suited theoretical tools
for describing medium and heavy nuclei. There are two main lines of
investigations based on the mean field approach, namely the
relativistic mean field (RMF) method where effective Lagrangians are
treated in the tree approximation, and the non-relativistic
Hartree-Fock (HF) method using effective interactions like the Gogny
and Skyrme forces. Recent reviews can be found in Refs.\cite{bender,
meng-toki}. When approaching the drip lines, one deals with open
shell nuclei where the effects of pairing correlations become quite
important, especially for such properties as the tails of matter
distributions. The pairing correlations can be described by the
non-relativistic Hartree-Fock-Bogoliubov (HFB) theory\cite{doba,gr},
or by the relativistic Hartree-Bogoliubov (RHB) theory\cite{meri0},
and one must add to the models a phenomenological pairing
interaction acting in the particle-particle channel.

Furthermore, the chemical potential $\lambda$ becomes close to zero
in the vicinity of drip lines and it is necessary to treat properly
the contributions of the quasiparticle continuum when evaluating the
pairing correlations\cite{gr}. Thus, the most appropriate approach
for such cases is to solve the self-consistent mean field equations
in coordinate space, and this is the method we use to obtain the
results of this work.


Important discrepancies are often found in the position of the
neutron drip line predicted by different models. For instance, these
differences clearly appear in the neutron drip line of Ni isotopes
calculated in relativistic\cite{meng1} and non-relativistic\cite{gr}
approaches. Even among different parametrizations of effective
Skyrme interactions used in HFB approach one finds different drip
line predictions, as illustrated in the recent work of
Ref.\cite{tera} and in the results of the present work. While
waiting for new experimental data which will help to discriminate
among the available models it can be interesting to analyze the
reasons for some of these discrepancies. This is one of the aims of
this paper.


A very interesting
phenomenon has been recently predicted within the RHB approach, the
formation  of a neutron giant halo (with up to six neutrons
involved) in some very neutron-rich isotopes. By plotting the
neutron distribution radius with respect to $A$ a kink appears where
the halo structure starts being formed. This effect has been found
to be particularly strong in Ca (with $A>60$) \cite{meng2} and Zr
isotopes (with $A>122$) \cite{meri}.
These predictions are based on the NLSH parametrization,
and similar results are obtained with the TM1
parametrization\cite{mengrev}. The giant halo phenomenon is also
found in the near-drip line Zr isotopes if one uses the NL3
parametrization in an RMF plus resonant continuum BCS\cite{sandu2}.
Until recently, the giant halo effect in medium-heavy nuclei has
been much less investigated within the non-relativistic mean field
approach, apart from Ref.\cite{tera} where the study of halo is
concentrated on the Ca isotopes. In this paper, we focus our
discussion on the Zr isotopes where we find that, in the HFB
approach with Skyrme forces the position of the neutron drip line
depends on the particular force adopted and consequently, the giant
halo effect seems to be more model dependent than in relativistic
approaches.

 A well known
phenomenon for which relativistic and non-relativistic predictions
are quite different is the charge isotope shift in Pb isotopes:
the kink in the charge radius at the shell closure $N=$ 126 is not
reproduced by non-relativistic calculations with the Gogny
interaction \cite{berger} or with standard Skyrme parametrizations
while it can be well reproduced by RMF calculations.
Reinhard and Flocard \cite{flore} have analyzed this discrepancy
with a special attention to the spin-orbit parametrization of Skyrme
interactions. New parametrizations
have been proposed in order to reproduce the observed kink of the
charge isotope shift.
If one introduces two parameters in the spin-orbit part of the
energy functional instead of one as in the standard Skyrme
interactions, it is possible
to obtain
a spin-orbit potential with isovector properties similar to that of
RMF. Then, a fitting procedure of all the Skyrme force parameters is
performed to produce new parametrizations which can describe well a
chosen set of nuclei and also reproduce the kink in the Pb isotope
shift. In the present calculations we have selected the
parametrization SkI4, and for comparison purposes we also use the
more standard set SLy4.

In this work we investigate the giant halo effect in Ca and Zr
isotopes within the non-relativistic Skyrme-Hartree-Fock-Bogoliubov
mean field approach. We find that the neutron drip line of Zr
isotopes calculated with SkI4 is far enough to give rise to a giant
halo phenomenon. On the other hand, neither SLy4 nor SkI4 can
produce bound Ca isotopes which are enough neutron-rich to lead to a
halo effect. The anti-halo effect due to pairing correlations
\cite{benna} is discussed for Zr and Ni isotopes by comparing HFB
results with the corresponding HF ones.


The article is organized as follows. In Sec. II we briefly describe
the theoretical framework. In Sec. III we present the calculated
two-neutron separation energies, neutron radii and single-particle
spectra for neutron-rich Ca and Zr isotopes and we compare our
non-relativistic results
with the corresponding RHB results obtained with the NLSH
parametrization \cite{meri,meng2}. In Sec. III we consider in more
detail the case of Zr isotopes and investigate giant halo and
anti-halo effects by analyzing occupation probabilities, neutron
density profiles and radii. The anti-halo effect is also studied for
Ni isotopes. In Sec. IV our conclusions are drawn.

\section{Theoretical framework}

The theoretical framework used in this paper is the
Hartree-Fock-Bogoliubov (HFB) approach. For local
two-body forces and spherical symmetry the HFB
equations have the following form:
\begin{equation}
\begin{split}
[ h ({\bf r}) -\lambda] u (E,{\bf r}) + \Delta ({\bf r})
v (E,{\bf r}) &= E u (E,{\bf r}) \,  \\
\Delta ({\bf r}) u (E,{\bf r}) - [h({\bf r})-\lambda] v (E,{\bf
r})&= E v(E,{\bf r}),
\end{split}
\label{eq02}
\end{equation}
where $\lambda$ is the Fermi energy, $h$ is the sum of the kinetic
energy and the HF mean field potential, and $\Delta$ is the pairing
potential; $u$ and $v$ are the upper and lower components of the
quasiparticle wave function associated with the energy $E$.

 The quasiparticle spectrum contains a discrete part for $E$ less than
 $-\lambda$ and a continuous part for $E$ above $-\lambda$.
 To calculate the continuum spectrum the HFB
 equations should be solved with  scattering type boundary conditions for
 the upper components of the HFB wave functions \cite{gr}. Since the continuum-HFB
 calculations are rather heavy, usually the continuum spectrum is discretized by
 imposing box boundary conditions, i.e., the condition that the HFB wave functions
 vanish at a given distance from the nucleus. This is sufficient for
 our present purpose provided that the box radius is properly
 chosen. We have checked that our results obtained with a box radius
 of 20 fm are very close to those of a full continuum calculation.

 In the HFB calculations presented in this paper the mean field is calculated with
a Skyrme type force while for the pairing channel we use a zero range interaction
with the following density dependence:
\begin{equation}
V({\bf r}_1 - {\bf r}_2) = V_0 \left[ 1- x \left(
\frac{\rho(r)}{\rho_0} \right) ^{\gamma} \right] \delta({\bf r}_1 -
{\bf r}_2). \label{e0}
\end{equation}
We have adopted the values $\rho_0$=0.16 fm$^{-3}$, $x$=0.5,
$\gamma$=1. The strength $V_0$ is
chosen so as to
reproduce the gaps extracted from the odd-even mass differences ( in
the regions where such experimental data are available).

As for the Skyrme interaction, we use two forces, i.e.,  SLy4 and
SkI4. The force SLy4, which includes constraints coming from the
neutron matter equation of state, is commonly employed for the
description of neutron-rich nuclei. The force SkI4 has a spin-orbit
potential similar to the RMF models and it includes in the fitting
protocol the requirement to reproduce the charge isotope shift for
Pb isotopes. Due to these characteristics one expects that this
Skyrme force will predict for the nuclei close to the drip line
properties similar to those of relativistic models.
We are particularly interested in the giant halo which might be
formed in heavy nuclei close to the neutron drip line. Neutron giant
halos were predicted by the RHB calculations in Ca and Zr regions
for $A>60$ and $A>122$, respectively \cite{meng1}. Thus, the pairing
strengths $V_0$ are adjusted to be -365 MeV $\cdot$ fm$^3$ and -290
MeV $\cdot$ fm$^3$ (-350 MeV $\cdot$ fm$^3$ and -300 MeV $\cdot$ fm$^3$) 
for the Ca and Zr
isotopes calculated with SLy4 (SkI4).

The Skyrme model predictions will be analyzed in the next section by
using the HFB approach. The HFB calculations are performed in
spherical symmetry, like in the RHB approach.

\section{ Results of HFB calculations}

\subsection{Separation energies and neutron radii}

The two-neutron separation energy is defined as
\begin{equation}
S_{2n} (N,Z)=E(N,Z)-E(N-2,Z),
\label{e1}
\end{equation}
where $E(N,Z)$ is the total energy of the isotope with $N$ neutrons
and $Z$ protons. The two-neutron separation energies  for Ca and Zr
isotopes are shown in Figs. 1 and 2. In these figures are displayed
only the separation energies for the bound nuclei (i.e., those
having a negative chemical potential in the HFB calculations). The
results correspond to the box-HFB calculations, which give
practically the same $S_{2n}$ values as continuum-HFB calculations.

The most important fact we can observe in Fig. 1 is the large
difference between the drip line location predicted by the
Skyrme-HFB with SLy4 or SkI4, and the RHB calculations of
Ref.\cite{meng2} for the Ca isotopes. It can be seen that, for both
Skyrme forces the drip line is located at $^{62}$Ca, while for RHB
calculations the drip line extends up to $^{72}$Ca. However, it is
found in Ref.\cite{tera} that the two-neutron separation energy is
still positive in $^{78}$Ca with the interaction SkM*. 
This strong sensitivity of the drip
line location to the particular Skyrme force makes difficult the
study of neutron-rich Ca isotopes.
In the Zr isotopes there are also differences among the model
predictions. As seen in Fig.2, for the force SLy4 the drip line is
located at $^{122}$Zr while for SkI4 at $^{138}$Zr. The latter
result is similar to the RHB one where the drip line is located at
$^{140}$Zr.

\begin{figure}
\begin{center}
\epsfig{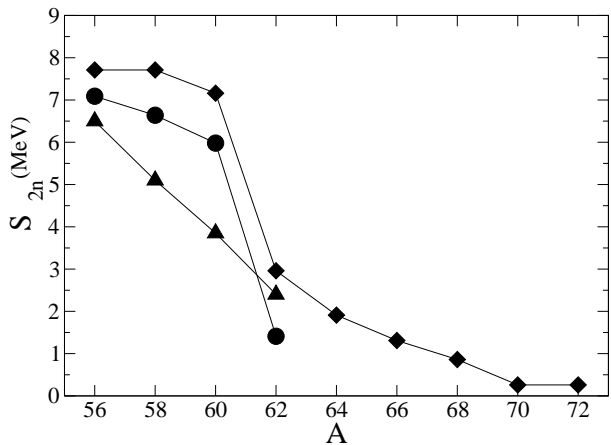}
\end{center}
\caption{Two-neutron separation energies for Ca isotopes calculated
with SLy4-HFB (triangles), SkI4-HFB (circles) and RHB (diamonds).
The RHB results correspond to Ref. \cite{meng2}.} \label{fig1}
\end{figure}

\begin{figure}
\begin{center}
\epsfig{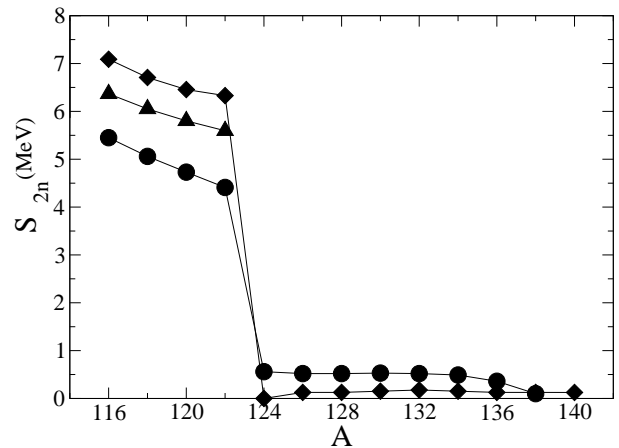}
\end{center}
\caption{Two-neutron separation energies for Zr isotopes. The
symbols are the same as in Fig. 1. The RHFB results correspond to
Ref.\cite{meri}.} \label{fig2}
\end{figure}
\begin{figure}
\begin{center}
\epsfig{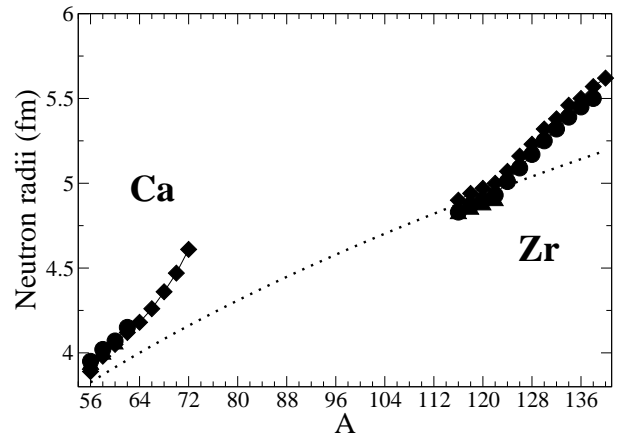}
\end{center}
\caption{Neutron radii for Ca  and Zr
isotopes. The symbols are the same as in Fig. 1 and the
RHB results correspond to Ref.\cite{meng2,meri}.
 Radii calculated with the formula $r_0 A^{1/3}$
($r_0=$ 1 fm) are shown by a dotted line. } \label{fig3}
\end{figure}
Next we analyze the neutron radii, shown in Fig. 3. As expected, for
Ca isotopes the HFB calculations with the interactions SLy4 and SKI4
do not predict a giant halo because the neutron drip line is reached
before the halo structure starts to be formed. The same is happening
for Zr isotopes if one uses the SLy4 force. However, for the force
SkI4 the situation is very different. Thus, as seen in Fig. 3, the
neutron radii given by the SkI4-HFB calculations show a change of
slope around $N=82$ and then follow the RHB results.

\begin{figure}
\begin{center}
\epsfig{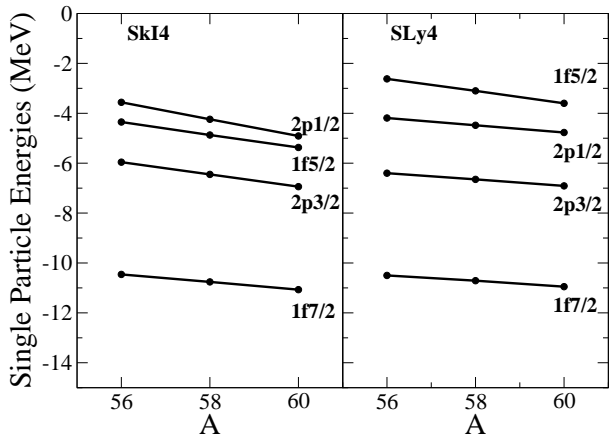}
\end{center}
\caption{SkI4-HF (left) and SLy4-HF (right) neutron single particle energies
for bound states in Ca
isotopes.}
\label{fig4}
\end{figure}
\begin{figure}
\begin{center}
\epsfig{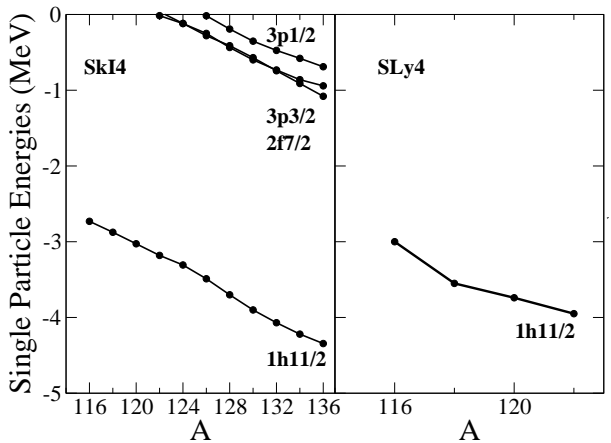}
\end{center}
\caption{SkI4-HF (left) and SLy4-HF (right) neutron single particle energies
for bound states in Zr
isotopes.}
\label{fig5}
\end{figure}

To understand better the behaviour of the radii, in Figs. 4 and 5
are shown the HF energies of the bound states close to the Fermi
level. For Ca isotopes we can see that, apart from a different
splitting of the $1f$ states, the two Skyrme forces give a rather
similar structure for the bound spectrum. In both cases the state
$1g9/2$ is not bound, at variance with the RMF calculations in which
this state becomes weakly bound at $A=62$ \cite{meng2}. Due to this
fact the drip line in the RHB calculations  is extended up to the
region where the giant halo can be formed, in contrast with the HFB
calculations of Ca isotopes based on SLy4 or SkI4. It must be noted,
however, that this tendency is not obeyed by the SkM*
parametrization which gives a bound $1g9/2$ orbital in the Ca
isotopes with A$\ge$60 resulting in a drip line located at higher
A\cite{tera}.

For Zr isotopes the structure of the bound spectrum is not the same
for the two Skyrme forces. We can observe that, for the force SLy4
there is only one bound state close to the Fermi level, i.e.,
$1h11/2$, while for the force SkI4 the states $2f7/2$, $3p3/2$ and
$3p1/2$ appear as bound states as well. These three weakly bound
states have a similar structure in the RMF calculations. This
explains why the SkI4-HFB and the RHB calculations provide a similar
halo in Zr isotopes. How these weakly bound HF states contribute to
the halo structure in the presence of pairing correlations is
discussed in the next subsection.

\subsection{The structure of the giant halo}

In the presence of pairing correlations the bound single-particle
states shown in Fig.5 are becoming quasiparticle resonances. To
describe them properly we have solved the HFB equations with
scattering type boundary conditions. In order to reduce the
numerical effort we have employed scattering boundary conditions
only for the quasiparticle states with the energy $-\lambda < E <
15$ MeV while for the other states we have used box boundary
conditions.

With the continuum HFB solutions we can calculate how the occupation
probability is changing in the region around a resonance. This
information is provided by the quantity
\begin{equation}
W(E)=\int_0^R dr r^2 v_E^2 (r),
\label{e2}
\end{equation}
where $v$ is the lower component of the HFB wave function and $R$ is
take equal to 20 fm. As an example, in Fig. 6 we show the values of
W(E) corresponding to the states  $3p1/2$, $3p3/2$ and $2f7/2$ in
the nucleus $^{132}$Zr. In this nucleus the quasiparticle continuum
starts at the energy $E=-\lambda=0.251$MeV.

\begin{figure}
\begin{center}
\epsfig{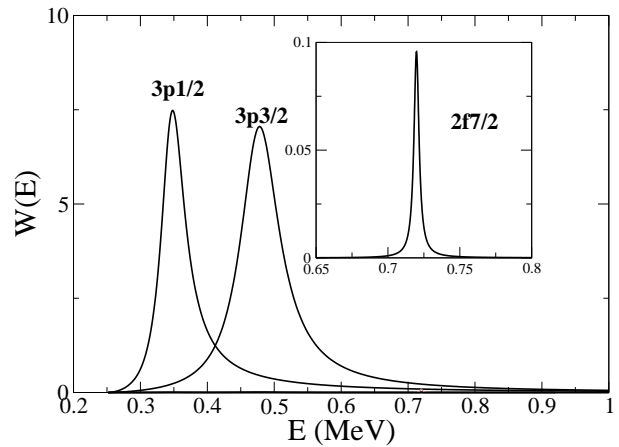}
\end{center}
\caption{Occupation profiles $W(E)$ (in units of MeV$^{-1}$) for the
states $3p1/2$, $3p3/2$ and $2f7/2$ in $^{132}$Zr, calculated with
SkI4-HFB.} \label{fig6}
\end{figure}
 Integrating the function W(E) over an energy interval in which it has
a significant value we can associate to each resonance an occupation
 probability $n$, i.e.,
\begin{equation}
n = \int_{E_1}^{E_2} W(E) dE
\label{e11}
\end{equation}
The occupation probabilities of the relevant resonant states in Zr isotopes
are shown in Fig. 7. It can be seen that the occupation probabilities
corresponding to the weakly bound states are increasing progressively
when passing from $^{124}$Zr to $^{138}$Zr. Thus these states contribute
significantly to the pairing correlations.

In order to analyze the structure of the halo, in Fig. 8 we plot the
quantity
\begin{equation}
R_{lj} (r)= \frac {\rho_{lj} (r)} {\rho(r)}
\label{e3}
\end{equation}
where $\rho$ and $\rho_{lj}$ are the total density and the density
corresponding to the channel ($l,j$), respectively. From Fig. 8 we
can clearly see that at large distances the dominant contribution to
the neutron density is given by the  $p$ states, which are less
confined by the centrifugal barrier compared to the other states
with higher $(l,j)$ values. This structure of the giant halo
obtained by using the SkI4-HFB model is very similar to that given
by the relativistic calculations \cite{meng1,sandu2}.

\begin{figure}
\begin{center}
\epsfig{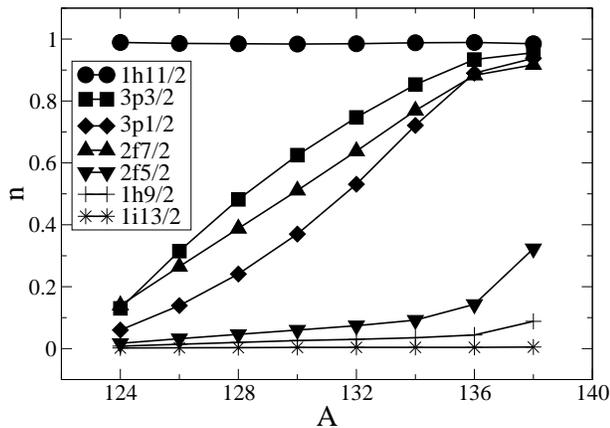}
\end{center}
\caption{Occupation probabilities for Zr isotopes, calculated with
SkI4-HFB.} \label{fig7}
\end{figure}

\begin{figure}
\begin{center}
\epsfig{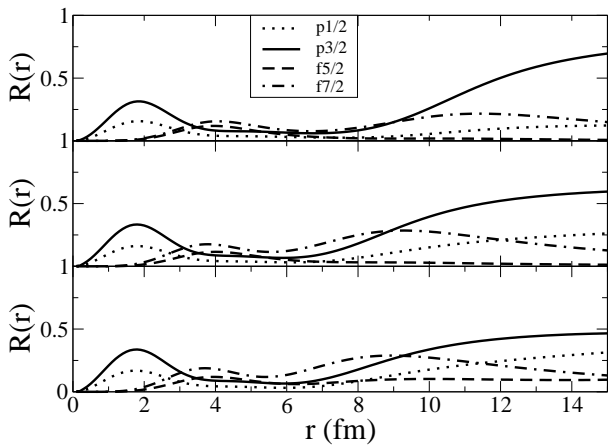}
\end{center}
\caption{Contributions of different $(l,j)$ channels
 to the total neutron density for
$^{124}$Zr (top), $^{132}$Zr (middle) and $^{138}$Zr (bottom). The
interaction is SkI4.} \label{fig8}
\end{figure}
\subsection{The anti-halo effect}

We turn now to the analysis of the so-called "anti-halo effect",
which was mainly discussed in light nuclei close to the neutron drip
line \cite{benna}. This effect is associated with a reduction of the
neutron radii by the pairing correlations. Here, we will show that
in fact the pairing correlations not only decrease but can also
increase the neutron radii in nuclei close to the drip line.

A strong effect of the pairing correlations upon the neutron radii is expected when
close to the Fermi level one finds weakly bound $s$ or $p$ states together with
states with higher angular momenta (weakly bound or resonances). As an illustration
we present below the case of Zr and Ni isotopes.

In Zr isotopes the pairing correlations affect the neutron halo
through the weakly bound $p$ and $f$ states. In Fig. 9 are shown the
neutron radii obtained in the HF and HFB calculations performed with
the force SkI4. We observe that in some cases ($A=$ 124, 128, 130, 136)
the HF radii are larger than the HFB ones while in other cases ($A=$
126, 132, 134) the contrary happens. This behaviour can be easily
understood if one considers the occupancy of the states $3p1/2$,
$3p3/2$ and $2f7/2$ in HF and HFB calculations shown in Table I and
Fig.7. We observe that for $A=124$ the state $3p3/2$ is half occupied
while $2f7/2$ is empty in HF. When pairing correlations are switched on,
the occupancy of  the state $3p3/2$ is reduced and that of the
$2f7/2$ is enhanced. Since the state $2f7/2$ has a smaller radial
extension than the state $3p3/2$ due to the centrifugal barrier, the
HFB radius is reduced compared to the HF radius. Similar
considerations apply for $A=$ 128 and 130. For $A=136$ the three states are 
completely occupied in HF. The HFB radius is thus smaller than the HF one due 
to the reduction of the occupancy of the two $p$ states when pairing is 
switched on. 

\begin{figure}
\begin{center}
\epsfig{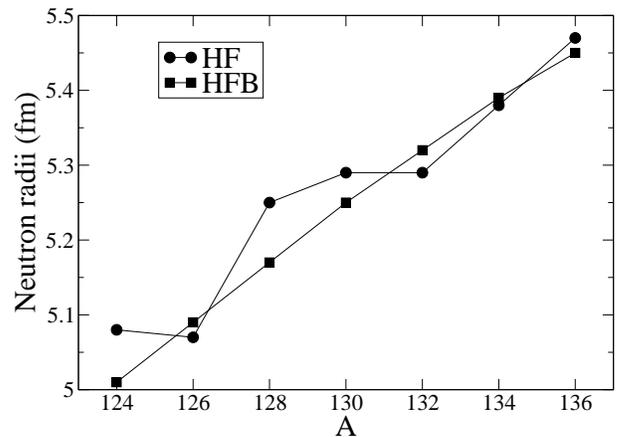}
\end{center}
\caption{HF (circles) and HFB (squares) neutron radii for Zr
isotopes. The interaction is SkI4.} \label{fig9}
\end{figure}

On the other hand, for $A=126$ the situation is opposite: in HF
approximation the state
$3p3/2$ is empty and the state $2f7/2$ is partially occupied.
Thus, in this case pairing  enhances the occupation of the state
$3p3/2$ and reduces that  of the state $2f7/2$. Consequently, the
radius becomes larger in HFB than in HF. A similar mechanism applies
for $A=132$ where the state $2f7/2$ is completely occupied
in HF and $3p3/2$ only partially occupied. For $A=134$ these two states
become completely occupied in HF. The pairing now scatters some
neutrons in the state $3p1/2$ which is empty in HF. The global effect
is a very weak enhancement of the neutron radius in HFB.
In conclusion, in Zr isotopes the pairing correlations can reduce or increase
the neutron radii according to the relative occupancy of the weakly bound
states close to the Fermi level.

\begin{table}
\begin{tabular}{|c|c|c|c|c|c|c|c|}
\hline
   $A$  & 124 & 126 & 128 & 130 & 132 & 134 & 136 \\
\hline
  $3p1/2$ & 0 & 0  & 0  & 0 & 0 & 0 & 2           \\
\hline
  $3p3/2$ & 2 & 0  & 4 & 4 & 2 & 4 & 4           \\
\hline
$2f7/2$ & 0 & 4 & 2 & 4 & 8 & 8 & 8          \\
\hline
\end{tabular}
\caption{\label{tabt} Number of neutrons in the states $3p1/2$, $3p3/2$ and
$2f7/2$ in HF as a function of $N$ in Zr isotopes.}
\end{table}

\begin{figure}
\begin{center}
\epsfig{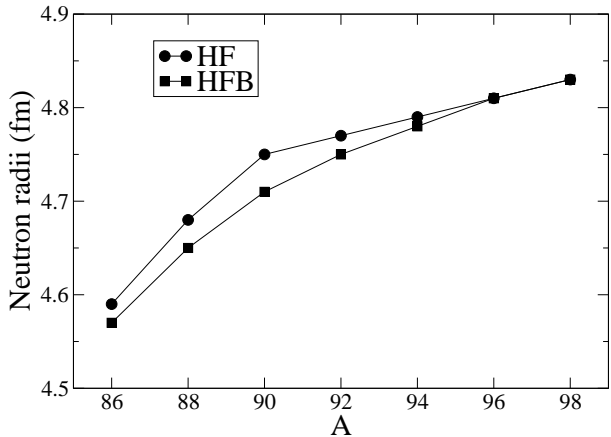}
\end{center}
\caption{HF (circles) and HFB (squares) neutron radii for Ni
isotopes. The interaction is SkI4.} \label{fig10}
\end{figure}

We complete our analysis by considering the Ni isotopes close to the
neutron drip line. The anti-halo effect in Ni isotopes was discussed
already in Ref.\cite{gr} by using the Skyrme force SIII. For this
force the neutron drip line is extended up to the nucleus $^{88}$Ni.
This prediction is very different from the RHB calculations, which
give $^{98}$Ni as the last bound nucleus. Interestingly enough, the
same position of the drip line, i.e.,$^{98}$Ni, is predicted by the
force SkI4.

The neutron radii of Ni isotopes calculated with the SkI4 force are
shown in Fig. 10. For the Ni isotopes analyzed here the most
important single-particle states are $3s1/2$, $2d3/2$ and $1g7/2$.
The first two states are weakly bound while $1g7/2$ is a resonance
state in $^{86}$Ni and $^{88}$Ni and becomes weakly bound beyond
$^{88}$Ni. This level structure is different from the case of Zr
isotopes in which all the states involved in the anti-halo effect
are bound states.

Coming back to the neutron radii shown in Fig. 10, we can see that,
from $^{86}$Ni to $^{94}$Ni the HF radii are larger than the HFB
ones, the difference being maximum for $^{90}$Ni. As in the case of
Zr isotopes, the effect of the pairing correlations upon neutron
radii ca be simply traced back to the occupancy of the HF levels
shown in Table II. The most important thing we can notice in Table
II is that the state $3s1/2$ is fully occupied in HF for all the
considered isotopes. Thus in all these nuclei the pairing force will
depopulate the state $3s1/2$, which has the largest spatial
extension compared to the other high-$l$ states shown in Table II.
This explains why in Ni isotopes the HF radii are systematically
larger than the HFB radii.

\begin{table}
\begin{tabular}{|c|c|c|c|c|c|c|c|}
\hline
   $A$  & 86 & 88 & 90 & 92 & 94 & 96 & 98 \\
\hline
  $3s1/2$ & 2 & 2  & 2  & 2 & 2 & 2 & 2           \\
\hline
  $2d3/2$ & 0 & 2  & 4 & 4 & 4 & 4 & 4           \\
\hline
$1g7/2$ & 0 & 0 & 0 & 2 & 4 & 6 & 8          \\
\hline
\end{tabular}
\caption{\label{tabt2} Number of neutrons in the states $3s1/2$,
$2d3/2$ and $1g7/2$ corresponding to Ni isotopes and HF calculations
with the SkI4 force.}
\end{table}

\vspace{0.4cm}
\section{Conclusions}

In this work we have examined the evolution of the nuclear structure
of Zr isotopes at large neutron excess, with a special attention to
the far out region of neutron densities. This is motivated by the
predictions of the RHB approach which indicate the presence of a
giant neutron halo in these nuclei. We have therefore used a
different approach, namely the Skyrme-HFB model. We find that, in
this isotopic chain the presence or absence of giant halos depends
essentially on the location of the predicted neutron drip line.
Thus, there is a strong model dependence in this type of study as
illustrated by the results obtained with the Skyrme forces SLy4 and
SkI4.

For the drip line to be displaced towards heavier isotopes, a
necessary condition is that some HF orbitals become bound when A
increases. In this case additional bound neutrons can be
accommodated and bound nuclei of heavier mass can be formed. An
illustration of this situation is provided by the  Zr isotopes where
this necessary condition is fulfilled by some model like SkI4 but
not by SLy4. Once the necessary condition is realized, a neutron
halo may exist if some of the weakly bound HF orbitals correspond to
low angular momenta (3p3/2 and 3p1/2 in the case of Zr with SkI4) so
that the centrifugal barrier is weak enough to let the wave
functions extend far out.

Thus, the decisive factor is the HF mean field while the pairing
correlations play a lesser role. Which part of the mean field
governs the position of the drip line is not clear, and this point
deserves further study. On the example of Zr with SkI4 and SLy4 it
seems that the main role cannot be attributed to the spin-orbit
potential but rather to some part of the mean field which influences
the relative positions of 1h and 3p orbitals.

We have also seen that the pairing correlations can lead to the
anti-halo effect, as it is well known, but it can also sometimes
produce an enhancement of the halo effect. This can be understood by
analyzing the occupation probabilities of the least bound (lj)
orbitals. Finally, it can be noted that the anti-halo effect can
occur even when all active orbitals are bound.

\end{document}